\documentclass[11pt,%
tightenlines,%
twoside,%
onecolumn,%
nofloats,
nobibnotes,%
nofootinbib,%
superscriptaddress,%
noshowpacs,%
centertags
]
{revtex4}
\usepackage{epsfig}
\usepackage{subfigure}
\usepackage{color}
\usepackage{xcolor}
\usepackage{amssymb}

\usepackage[centertags]{amsmath}
\usepackage{amsfonts}
\usepackage{amsthm} 
\usepackage{newlfont}
\usepackage{epsfig}
\usepackage{amscd}
\usepackage{footnote}
\usepackage{lipsum}
\usepackage{subcaption}
\usepackage{multirow}
\usepackage{amscd}

\newcommand{\beq}{\begin{equation}}
\newcommand{\eeq}{\end{equation}}
\newcommand{\ba}{\begin{array}}
\newcommand{\ea}{\end{array}}
\newcommand{\bea}{\begin{eqnarray}}
\newcommand{\eea}{\end{eqnarray}}

\begin{document}

\title{Some Aspects of Remote State Restoring in~State Transfer Governed by XXZ-Hamiltonian}

\author{\firstname{Georgii A.}~\surname{Bochkin}}
\affiliation{Federal Research Center of Problems of Chemical Physics
and Medicinal Chemistry, Russian Academy of Sciences, Chernogolovka,
Ac. Semenov ul. 1, Moscow reg. 142432, Russia}

\author{\firstname{Sergei I.}~\surname{Doronin}}
\affiliation{Federal Research Center of Problems of Chemical Physics
and Medicinal Chemistry, Russian Academy of Sciences, Chernogolovka,
Ac. Semenov ul. 1, Moscow reg. 142432, Russia}

\author{\firstname{Edward  B.}~\surname{Fel'dman}}
\email[E-mail: ]{efeldman@icp.ac.ru} \affiliation{Federal Research
Center of Problems of Chemical Physics and Medicinal Chemistry,
Russian Academy of Sciences, Chernogolovka, Ac. Semenov ul. 1,
Moscow reg. 142432, Russia}

\author{\firstname{Elena I.}~\surname{Kuznetsova}}
\affiliation{Federal Research Center of Problems of Chemical Physics
and Medicinal Chemistry, Russian Academy of Sciences, Chernogolovka,
Ac. Semenov ul. 1, Moscow reg. 142432, Russia}

\author{\firstname{Ilia D.}~\surname{Lazarev}}
\affiliation{Federal Research Center of Problems of Chemical Physics
and Medicinal Chemistry, Russian Academy of Sciences, Chernogolovka,
Ac. Semenov ul. 1, Moscow reg. 142432, Russia}

\author{\firstname{Alexander}~\surname{Pechen}}
\email[E-mail: ]{apechen@gmail.com} \affiliation{Department of
Mathematical Methods for Quantum Technologies, Steklov Mathematical
Institute of Russian Academy of Sciences, Gubkina ul. 8, Moscow
119991, Russia} \affiliation{Quantum Engineering Research and
Education Center, University of Science and Technology MISIS,
Leninskiy prospekt 6, Moscow, 119991, Russia}

\author{\firstname{Alexander I.}~\surname{Zenchuk}}
\email[E-mail: ]{zenchuk@itp.ac.ru} \affiliation{Federal Research
Center of Problems of Chemical Physics and Medicinal Chemistry,
Russian Academy of Sciences, Chernogolovka, Ac. Semenov ul. 1,
Moscow reg. 142432, Russia}



\begin{abstract}
We consider the remote state restoring and perfect transfer of  the
zero-order coherence matrix (PTZ) in a spin system governed by the
XXZ-Hamiltonian conserving the excitation number. The restoring tool
is represented by several nonzero Larmor frequencies in the
Hamiltonian. To simplify the analysis we use two approximating
models including either step-wise or  pulse-type time-dependence of
the Larmor frequencies. Restoring in spin chains with up to 20 nodes
is studied. Studying  PTZ, we consider the zigzag and rectangular
configurations and optimize the transfer of the 0-order coherence
matrix  using geometrical parameters of the communication line as
well as the special unitary transformation of the  extended
receiver.  Overall observation is that XXZ-chains require longer
time for state transfer than XX-chains, which is confirmed by the
analytical study of the evolution under the nearest-neighbor
approximation. We  demonstrate the exponential increase of the
state-transfer time  with the spin chain length.
\end{abstract}



\maketitle

\section{Introduction}

The problem of information exchange  between two remote subsystems
(conventionally called sender and receiver) is an attractive and
important area of quantum information \cite{Bose,CDEL,KS,GKMT}. The
remote state restoring \cite{Z_2018,BFLPZ_2022} is a method of
information transfer from the sender to the receiver with minimal,
in certain sense, deformation. By the minimal deformation we mean
obtaining such a state that (almost) each element of the receiver's
density matrix is proportional to the appropriate element of the
sender's density matrix. We say ``almost'' because there is an
obstacle for restoring  all diagonal elements, which arises from the
trace-normalization condition for the density matrix. Although this
problem was overcome in the protocol presented in
\cite{BFLPZ_2022}, this protocol unavoidably imposes certain
constraint on the structure of the state to be transferred. Namely,
some set of elements of the sender's density matrix must be zero. It
is clear that the state-restoring technique is very close to the
method of controlling quantum states developed in papers
\cite{AubourgJPB2016,ShanSciRep2018,PyshkinNJP2018,FerronPS2022,PH_2006,PISR_2006}.

Originally, a unitary transformation of the extended receiver was
proposed as a  state-restoring tool \cite{Z_2018}. However, the
multiqubit  unitary transformation is hard to implement for an
effective realization \cite{KShV,NCh}. Therefore, an alternative
generic approach based on using a suitably optimized inhomogeneous
magnetic field was proposed and developed in \cite{FPZ_Archive2023},
where general formulas which can be applied to various Hamiltonians
including XXZ models  were derived and applied to the XX
Hamiltonian, which is a particular case of XXZ model. In this case,
inhomogeneous time-dependent magnetic field is introduced through
the step-wise (and pulse-type) time-dependent Larmor
frequencies~\cite{PP_2023,MP_2023,FPZ_Archive2023} which can be
called the controlling Larmor frequencies.  It was shown for the 2-
and 3-qubit sender/receiver that the  minimal number of the
controlling Larmor frequencies equals, respectively, 2 and 3.

In this paper, we consider several  aspects associated with the
remote state restoring of the state transferred from the sender to
the receiver under XXZ-Hamiltonian with the sender and receiver of
the same dimensionality: $N^{(R)}=N^{(S)}$. As a state restoring
tool, we use the parameters of the external  inhomogeneous magnetic
field (several nonzero Larmor frequencies in the Hamiltonian), which
optimize the state-restoring protocol. Constructing state-restoring
protocol we deal with (0,1)-excitation state
subspace~\cite{FPZ_Archive2023}.

Since restoring  diagonal elements  does not obey the general rule
of restoring  non-diagonal elements, we concentrate on perfect
transfer of the zero-order  coherence matrix (PTZ) and perform
optimization of  this process via geometric parameters of
2-dimensional models such as zigzag  and rectangular (multichannel)
communication lines. In this process, we consider the state subspace
involving  states with 0-, 1- , 2-, and 3-excitations.

As was shown in \cite{PRA2010},  the evolution under XXZ-Hamiltonian
requires longer time for the state transfer. We demonstrate via the
analytical formulas proposed in Ref.~\cite{PRA2010} for the 2-qubit
state transfer that the optimal time instant for  the receiver's
state registration exponentially increases with the spin chain
length.

The paper is organized as follows.  In Sec.~\ref{Section:MagF}, we
consider the protocol for remote restoring states with
(0,1)-excitations via the parameters of inhomogeneous magnetic
field. Perfect transfer of 0-order coherence matrix via zigzag and
rectangular spin configurations is studied in
Sec.~\ref{Section:0cm}. Some  aspects of the analytical approach to
the remote state restoring are considered in
Sec.~\ref{Section:analyt}. Concluding remarks are given in
Sec.~\ref{Section:Conclusions}.

\section{Remote state restoring in a linear line via an inhomogeneous magnetic field}
\label{Section:MagF}

Governed by some Hamiltonian $H$,  evolution of a density matrix
$\rho$ is described by the Liouville--von Neumann equation (where we
set the Planck constant $\hbar=1$)
\begin{eqnarray}\label{rho}
i \rho_t(t) = [H(t),\rho(t)]=H(t)\rho(t) - \rho(t) H(t)
\end{eqnarray}
with some initial state  $\rho(0)$.

We deal with a one-dimensional communication  line including the
sender $S$ with $N^{(S)}$ nodes, receiver $R$ with
$N^{(R)}=N^{(S)}$ nodes and transmission line $TL$  with
$N^{(TL)}$ nodes connecting them.  We  solve the initial value
problem with the initial state
\begin{eqnarray}\label{rho0}
\rho(0)=\rho^{(S)}(0)\otimes \rho^{(TL;R)}(0),
\end{eqnarray}
where $\rho^S(0)$ is an arbitrary  initial sender's state to  be
transferred to the receiver $R$, while the initial state of the
transmission line and receiver $\rho^{(TL;R)}(0)$ is the ground
state,
\begin{eqnarray}\label{TLR}
\rho^{(TL;R)}(0) =  {\mbox{diag}}(1,0,\dots).
\end{eqnarray}
We consider evolution under the time-dependent XXZ-Hamiltonian \cite{Abragam}
\begin{eqnarray}\label{HH0}
H(t)&=&H_0 +\Omega(t),\quad \Omega(t)= \sum_{j=N-N^{(\Omega)}+1}^N
\omega_j(t) Z_{j},\quad \omega_j =\gamma B_j,
\end{eqnarray}
where
\begin{eqnarray}
\label{XX}
H_0&=&\sum_{j>i} D_{ij} (X_{i}X_{j}+Y_{i}Y_{j}  - 2 Z_i Z_j ),\quad D_{ij}=\frac{\gamma^2 }{r^3_{ij}},\\
X_i&=&\frac{1}{2}\left(\begin{array}{cc}
0&1\cr
1&0
\end{array}\right),\quad Y_i=\frac{1}{2}\left(\begin{array}{cc}
0&-i\cr
i&0
\end{array}\right),\quad Z_i=\frac{1}{2}\left(\begin{array}{cc}
1&0\cr
0&-1
\end{array}\right),\nonumber
\end{eqnarray}
and
\begin{equation}
[H_0,I_z]=0.\label{com}
\end{equation}
Here $\gamma$ is the  gyromagnetic   ratio, $r_{ij}$ is the distance
between the $i$th and $j$th spins and we set $\hbar=1$. The
time-dependence is represented by $N^{(\Omega)}$ Larmor frequencies
$\omega_j$.

The commutation relation (\ref{com})  prescribes the block-diagonal
structure to the Hamiltonian written in the basis sorted by
excitation number
\begin{eqnarray}\label{bH}
H={\mbox{diag}} (H^{(0)},H^{(1)},\dots ),
\end{eqnarray}
where the  block $H^{(j)}$ governs the  evolution of the
$j$-excitation subspace and $H^{(0)}$ is a scalar:
\begin{eqnarray}
\nonumber
H^{(0)} =H_0^{(0)}+ \sum_{i=N-N^{(ER)}+1}^N \omega_i.
\end{eqnarray}
Since there is a freedom in the energy of the ground  state we can
require $H^{(0)}=H_0^{(0)}$ that yields the constraint on the
control functions $\omega_i$,
\begin{eqnarray}\label{omega2}
\sum_{i=N-N^{(\Omega)}+1}^N \omega_i =0.
\end{eqnarray}
Hereafter we use dimensionless time $\tau$,
\begin{eqnarray}\label{tau}
\tau=t D_{12},
\end{eqnarray}
and, following Ref.~\cite{FPZ_Archive2023},  consider two models
fixing the time-dependence of the Larmor frequencies.  For both
models, we study the evolution over the time interval $0\le \tau \le
\tau_{\mathrm{reg}}$, where $\tau_{\mathrm{reg}}$ is the time
instant selected for the receiver's state registration and will be
specified below.

{\bf Model~1.} Let us consider the step-wise $\tau$-dependence of
the Larmor frequencies
\begin{eqnarray}\label{LF}
\omega_k(\tau) = \sum_{j=1}^{K_\omega} a_{kj} \theta_j(\tau),\quad
\theta_j(\tau) =\left\{\begin{array}{ll} 1, &\tau_{j-1} < \tau\le
\tau_j,\cr 0,  & {\mbox{otherwise},} \end{array} \right. \;\; k=
N-N^{(\Omega)}+1, \dots, N.
\end{eqnarray}
Here we split the entire time interval $[0,\tau_{\mathrm{reg}}]$ in
$K_\omega$ equal intervals, $\Delta \tau =
\tau_{j}-\tau_{j-1}={\tau_{\mathrm{reg}}}/{K_\omega}$. Constraint
(\ref{omega2}) results in the corresponding constraints on the
control parameters $a_{kj}$:
\begin{eqnarray}
\nonumber
\sum_{k=N-N^{(\Omega)}+1}^{N} a_{kj}=0,\;\;1\le j\le K_\omega.
\end{eqnarray}
We set
\begin{eqnarray}
\nonumber
a_{Nj}=-\sum_{k=N-N^{(ER)}+1}^{N-1} a_{kj},\;\;1\le j\le K_\omega.
\end{eqnarray}
Therefore, we have $K_\omega(N^{(\Omega)}-1)$ free parameters.

Thus, the Hamiltonian is piecewise constant, therefore, we can
analytically integrate the Liouville equation (\ref{rho}) over each
interval $\Delta\tau$ and the whole evolution operator  over the
interval $[0,\tau_{\mathrm{reg}}]$ can be represented by the
product of the operators
\begin{eqnarray}\label{VV}
U_j(\tau-\tau_{j-1}) &=& \left\{
\begin{array}{ll}
\exp(-i \frac{H_j}{D_{12}} (\tau-\tau_{j-1})),& \tau_{j-1} < \tau\le \tau_j,\cr
1, & {\mbox{otherwise},}
\end{array}
\right.\\
\nonumber && H_j = H_0 +\Omega_j,\quad   \Omega_j=\sum_i a_{ij}
I_{zi},\quad j=1,\dots,K_\omega, \label{Omega}
\end{eqnarray}
as follows
\begin{eqnarray}\label{UtK}
U(\tau_{\mathrm{reg}})=  U_{K_\omega}(\Delta \tau
)\dots U_1(\Delta \tau) U(0).
\end{eqnarray}
We approximate  the evolution operator using the Trotterization method \cite{Trotter,Suzuki},
\begin{eqnarray}\label{Vtr}
U_j(\Delta\tau) = e^{-i \frac{H_j}{D_{12}} \Delta \tau}\approx
U^{(n)}_j (\Delta\tau) = \left(e^{-i
\frac{{H_0}}{D_{12}}\frac{\Delta\tau}{n}} e^{-i
\frac{\Omega_j}{D_{12}}\frac{\Delta\tau}{n}}\right)^n
\end{eqnarray}
(where $n$ is the Trotterization number) and replace the operator  $U$  by $U^{(n)}$ in (\ref{UtK}),
\begin{eqnarray}
\nonumber
U^{(n)}(\tau_{\mathrm{reg}})=\prod_{j=1}^{K_\omega} \left(U_j\Big(\frac{\Delta\tau}{n}\Big)\right)^{n}=
\prod_{j=1}^{K_\omega} \left(U_j\Big(\frac{\tau_{\mathrm{reg}}}{K_\omega n}\Big)\right)^{n}.
\end{eqnarray}

{\bf Model~2.} Similar to the Model~1, we split the  interval $\tau$
into the set of $K_\omega$ subintervals $\Delta \tau$. But now, in
addition, we split each $\Delta \tau$ into 2 subintervals, $ \Delta
\tau= \Delta \tau^{(1)}+ \Delta \tau^{(2)}, $ so that
\begin{eqnarray}\label{HHj}
H_j=\left\{ \begin{array}{ll} H_0, & \tau \in \Delta\tau^{(1)},\cr
H_0 + \Omega_j, & \tau \in \Delta\tau^{(2)}.
\end{array}
\right.
\end{eqnarray}
Thus, for this model,  we consider the $\tau$-dependence of  the
Larmor frequencies $\omega_j$  in the pulse form. We require
\cite{FPZ_Archive2023}
\begin{eqnarray}
\label{avar}
\frac{\|\Omega_j\|}{D_{12}} \Delta\tau^{(2)} \sim \frac{\|H_0\|}{D_{12}} \Delta\tau^{(1)}  ,
\end{eqnarray}
where $\|\cdot\|$ is some matrix norm. Large
$\|\Omega_j\|$s act over the short time intervals. This means that
the term $\Omega_j$  in the sum $H_j=H_0+\Omega_j$ (see
Eq.~(\ref{VV})) dominates over $H_0$ on the interval
$\Delta\tau^{(2)}$. Therefore, we can approximate
Hamiltonian~(\ref{HHj}) as
\begin{eqnarray}\label{HHj2}
H_j\approx \left\{ \begin{array}{ll} H_0, & \tau \in
\Delta\tau^{(1)},\cr \Omega_j, & \tau \in \Delta\tau^{(2)},
\end{array}
\right.
\end{eqnarray}
and write the approximate evolution operator $U_j(\Delta \tau)$ as
\begin{eqnarray}
\nonumber
U_j(\Delta \tau)=    e^{-i \frac{H_0}{D_{12}} \Delta\tau^{(1)}}  e^{-i \frac{H_0+\Omega_j}{D_{12}} \Delta\tau^{(2)}} \approx
  e^{-i \frac{H_0}{D_{12}} \Delta\tau^{(1)}}e^{-i \frac{\Omega_j}{D_{12}} \Delta\tau^{(2)}}.
\end{eqnarray}
We denote the evolution operator by $U^{(0)}$ to distinguish it from
the Trotterized  operator $U^{(n)}$, $n>0$. We also introduce the
parameter $\varepsilon$,
\begin{eqnarray}\label{var0}
\varepsilon={\Delta\tau^{(2)}}/{\Delta\tau},
\end{eqnarray}
that characterizes the relative duration of $\Delta \tau^{(2)}$. Thus, we have
\begin{eqnarray}
\nonumber U^{(0)}(\tau_{\mathrm{reg}},\varepsilon) =
\prod_{j=1}^{K_\omega}
U_{j}\left(\frac{\tau_{\mathrm{reg}}}{K_\omega},\varepsilon\right) .
\end{eqnarray}

\subsection{Restoring of (0,1)-excitation state}
The receiver's density matrix $\rho^{(R)}$ is obtained  from $\rho$
by taking the partial trace
\begin{eqnarray}\label{R}
\rho^{(R)}={\mbox{Tr}}_{S,TL}(\rho).
\end{eqnarray}
State restoring means creating such  $\rho^{(R)}$ that
\begin{eqnarray}\label{rest}
\rho^{(R)}_{ij}(t_{\mathrm{reg}}) = \lambda_{ij} \rho^{(S)}_{ij}(0), \quad i\neq j,
\end{eqnarray}
which holds for all nondiagonal elements of $\rho^{(R)}_{ij}$.   The
restoring of the diagonal elements is nontrivial because of the
normalization ${\mbox{Tr}}\rho^{(R)} =1$. However, it was shown in
\cite{TZarxiv} that almost all diagonal elements can be restored in
the case of (0,1)-excitation states except $\rho^{(R)}_{11}$
(probability for the ground state of the receiver). We shall
emphasize that the scale parameters $\lambda_{ij}$ are universal,
i.e., they do not depend on the element of  the initial sender's
density matrix and are completely defined by the Hamiltonian and the
selected time instant for state registration $t_{\mathrm{reg}}$.

Let us write condition (\ref{rest}) in terms of the  operator $U$.
To this end, it is convenient to introduce the multi-index notation
\begin{eqnarray}\label{MI}
I=\{I_S I_{TL} I_R\},
\end{eqnarray}
where each subscript indicates the appropriate subsystem:  sender
$S$, transmission line $TL$ and receiver $R$.  Each multi-index
$I_S$, $I_{TL}$ and $I_R$ is a sequence of binary digits whose
length equals the number of spins in the appropriate subsystem, each
digit takes value either 1 (excited spin) or 0 (spin in the ground
state). By $|I|$, $|I_S|$, etc. we denote the cardinality of the
corresponding set of indices, i.e., the number of units in the
corresponding multi-index. Then, Eq.~(\ref{R}) can be written as
follows
\begin{eqnarray}\label{rhoR}
\rho^{(R)}_{N_R;M_R} = \sum_{I_S,J_S} T_{N_RM_R;I_SJ_S}\rho^{(S)}_{I_S;J_S} ,
\end{eqnarray}
where the matrix elements of $T$ are expressed  in terms of the
matrix elements of $U$ as follows
\begin{eqnarray}\label{TT}
  T_{N_RM_R;I_SJ_S}&\equiv&\sum_{{N_S, N_{TL}}\atop{ |N_S|+|N_{TL}|=   |I_S|-|N_R|=|J_S|-|M_R|}}
  U_{N_SN_{TL}N_R; I_S0_{TL}0_R} U^\dagger_{J_S 0_{TL}0_R; N_SN_{TL}M_R}.
\end{eqnarray}
The tensor $T$ transfers the initial   density matrix of the sender
$S$  to the receiver $R$  and is represented by a composition of
Kraus operators \cite{KrausBook}. Writing the sum in (\ref{TT}), we
take into account the constraint $|N|=|I|$ for subscripts in
$U_{N;I}$ that  is required by the commutation relation (\ref{com}).
In addition, $ U_{0_S0_{TL}0_R; 0_S0_{TL}0_R}=U_0 $ is a scalar.

Next, we have to satisfy the restoring constraint (\ref{rest})  for
the non-diagonal elements of $\rho^{(R)}$  which yields
\begin{eqnarray}
\nonumber \lambda_{N_R0_R} &=&T_{N_R0_R;N_R0_R}= T_{0_RN_R;0_RN_R}^*
=U_{0_S0_{TL} N_R; N_R 0_{TL} 0_R}U_0^* ,\\\nonumber
\lambda_{N_RM_R}&=&T_{N_RM_R;N_RM_R} = U_{0_S0_{TL}N_R;
N_R0_{TL}0_R} U^\dagger_{M_R 0_{TL}0_R; 0_S0_{TL}M_R}\\\nonumber
&&=\lambda_{N_R0_R}\lambda_{0_RM_R} =
 \lambda_{N_R0_R}\lambda^*_{M_R0_R},\\
\nonumber
T_{N_RM_R;I_SJ_S}
 &=& U_{0_S0_{TL}N_R; I_S0_{TL}0_R} U^\dagger_{J_S 0_{TL}0_R; 0_S0_{TL}M_R} =0,   \\\nonumber
 &&\;\;N_R\neq I_S,\quad M_R\neq J_S, \quad N_R\neq M_R,
\end{eqnarray}
which is equivalent to
\begin{eqnarray}\label{constr_red}
 U_{0_S0_{TL}N_R; I_S0_{TL}0_R} =0,\quad |I_S| =|N_R| =1,\quad N_R\neq I_S.
\end{eqnarray}
In addition, for the diagonal elements of  $\rho^{(R)}$,  we have
representation (\ref{rhoR}) with $N_R=M_R$ and
\begin{eqnarray}
\nonumber
T_{N_RN_R;I_SJ_S}
 &=& U_{0_S0_{TL}N_R; I_S0_{TL}0_R} U^\dagger_{J_S 0_{TL}0_R; 0_S0_{TL}N_R} ,\quad N_R\neq 0_R,
\end{eqnarray}
so that
\begin{eqnarray*}
\nonumber T_{N_RN_R;I_SJ_S} =0 \;\;{\mbox{if}} \;\; N_R\neq I_S
\;\;{\mbox{or}}\;\; N_R\neq J_S,\quad T_{N_RN_R;N_RN_R} =
|\lambda_{N_R 0_R}|^2.\nonumber
\end{eqnarray*}
Moreover,
\begin{eqnarray}\nonumber
T^{(n)}_{0_R0_R;I_SJ_S} &=& \sum_{{N_S,
N_{TL}}\atop{|N_S|+|N_{TL}|=|I_S|=|J_S|}} U_{N_SN_{TL}0_R; I_S0_{TL}0_R}
(U^{(n)})^\dagger_{J_S 0_{TL}0_R;
N_SN_{TL}0_R}+\delta_{I_S0_S}\delta_{J_S0_S}.
\end{eqnarray}
Thus, according to (\ref{rhoR}), the diagonal elements   of
$\rho^{(R)}$ are all restored except for the element
$\rho^{(R)}_{0_R0_R}$.

Now we remember that the evolution is approximated using  two Models
above. Therefore, in numerical simulations, we replace the operator
$U$ by $U^{(n)}$, the $\lambda$-parameters by the
$\lambda^{(n)}$-parameters, the set of Larmor frequencies $\omega$
by $\omega^{(n)}$. Here $n=0$ means Model~2, while $n>0$ means the
Trotterization number in Model~1.  In particular,
Eq.~(\ref{constr_red}) becomes
\begin{eqnarray}\label{constr_red_appr}
 U^{(n)}_{0_S0_{TL}N_R; I_S0_{TL}0_R} (\omega^{(n)}) =0,\quad |I_S| =|N_R| =1,\quad N_R\neq I_S.
\end{eqnarray}
Solution $\omega^{(n)}$ of this equation is not  unique. We denote
different solutions by $\omega^{(n;m)}$, $m=1,2,\dots$. Since
$\omega^{(n;m)}$ are obtained using approximate equation
(\ref{constr_red_appr}), then
\begin{eqnarray}
\nonumber
U_{0_S0_{TL}N_R; I_S0_{TL}0_R} (\omega^{(n)}) \neq 0,
\quad |I_S| =|N_R| =1,\quad N_R\neq I_S.
\end{eqnarray}
Now we have to estimate the accuracy of Models~1 and 2 introducing
the function
\begin{eqnarray}\label{epsilon}
&&
S^{(n)}_1(\tau_{\mathrm{reg}}) =\min_m \max_{N_R\neq I_S}
U_{0_S0_{TL}N_R; I_S0_{TL}0_R} (\omega^{(n;m)},\tau_{\mathrm{reg}}),\quad n\ge 0.
\end{eqnarray}
We emphasize that formula (\ref{epsilon}) involves $U$ rather then $U^{(n)}$.

Another function is associated with the accuracy of constructing the $\lambda$-parameters, i.e.,
\begin{eqnarray}\label{epsilon2}
S^{(n)}_2(\tau_{\mathrm{reg}}) =\min_m\max_{N_R} \left|\lambda^{(n)}_{N_R0}(\omega^{(n,m)},\tau_{\mathrm{reg}})) -
\lambda_{N_R 0}(\omega^{(n,m)},\tau_{\mathrm{reg}}))\right|,
\end{eqnarray}
which is also zero in the case of perfect approximation.  The
following two characteristics associated with $S_1^{(n)}$ and
$S_2^{(n)}$ can be also useful:
\begin{eqnarray}\label{S3}
&&
S^{(n)}_3(\tau_{\mathrm{reg}}) = \max_{0\le\tilde \tau\le \tau_{\mathrm{reg}}} S^{(n)}_1(\tilde \tau),
\\\label{S4}
&&
S^{(n)}_4(\tau_{\mathrm{reg}}) = \max_{0\le\tilde \tau\le \tau_{\mathrm{reg}}} S^{(n)}_2(\tilde \tau).
\end{eqnarray}
We emphasize that the functions $S^{(n)}_1(\tau_{\mathrm{reg}})$
and $S^{(n)}_2(\tau_{\mathrm{reg}})$ indicate the local (in time)
accuracies of solving system (\ref{constr_red_appr}) and calculating
$\lambda$-factors.  On the contrary,
$S^{(n)}_3(\tau_{\mathrm{reg}})$ and
$S^{(n)}_4(\tau_{\mathrm{reg}})$ are the global characteristics of
a model informing  what are the best accuracies  $S_1^{(n)}$ and
$S_2^{(n)}$ over the interval $[0,\tau_{\mathrm{reg}}]$ at fixed
chain length $N$ and sender/receiver dimension $N^{(S)}$. Now we
introduce the function $S^{(n)}_5(\tau_{\mathrm{reg}})$,
\begin{eqnarray}\label{S5}
S^{(n)}_5(\tau_{\mathrm{reg}}) = \max_{0\le\tilde \tau\le
\tau_{\mathrm{reg}}} \max_{m}  \min_{N_R} \left|\lambda
^{(n)}_{N_R0} (\omega^{(n;m)},\tilde \tau)\right|,
\end{eqnarray}
which shows how accurately we have restored  the transferred state;
$S_5^{(n)}=1$ in the case of perfect state transfer. Of course,
$S^{(n)}_5$ depends on the chain length $N$, which is another
argument of this  function, $S^{(n)}_5(\tau,N)$. The function
$S^{(n)}_5(T(N),N)$ with some fixed large enough time-interval $T$
depending on $N$ characterizes the effectiveness of application of
state-restoring protocol to the chains of different lengths over the
time interval $T$.  With $S^{(n)}_5(T,N)$, we associate the time
instant for  state registration  $\tau_{\mathrm{reg}} =
\,\tau_0(N)$, $0<\tau_0(N)< T $, such that
\begin{eqnarray}\label{lamn2}
\lambda^{(n)}=S^{(n)}_5(T,N) = S^{(n)}_5(\tau_0(N) ,N).
\end{eqnarray}
In other words, $\tau_0$ is the time instance inside  of the above
interval $T$ at which the state restoring provides the best
optimization result.

\subsection{Numerical simulations}
\label{Section:Num} In all examples of this section we deal with
(0,1)-excitation evolution.  We set $N^{(S)} = N^{(R)}$ and consider
a homogeneous chain with sender and receiver of  2 or 3 spins. To simplify the control of
boundedness of $|a_{ij}|$ obtained as  numerical solutions of system
(\ref{constr_red}), we represent $a_{ij}$ as
\begin{eqnarray}\label{ta}
a_{ij} =2 \sin \tilde a_{ij}.
\end{eqnarray}
Thus, $-2\le a_{ij}\le 2$.

First, we give a detailed study of the short  chain $N=6$ and show
that both  approximating models, Model~1 and Model~2, are quite
reasonable at, respectively, large enough Trotterization number $n$
and small parameter $\varepsilon$.

\subsubsection{Two-qubit sender $N^{(S)}=N^{(R)}=2$ in spin chain of $N=6$ nodes.}
\begin{figure*}[!]
\centering
\includegraphics[width=\textwidth]{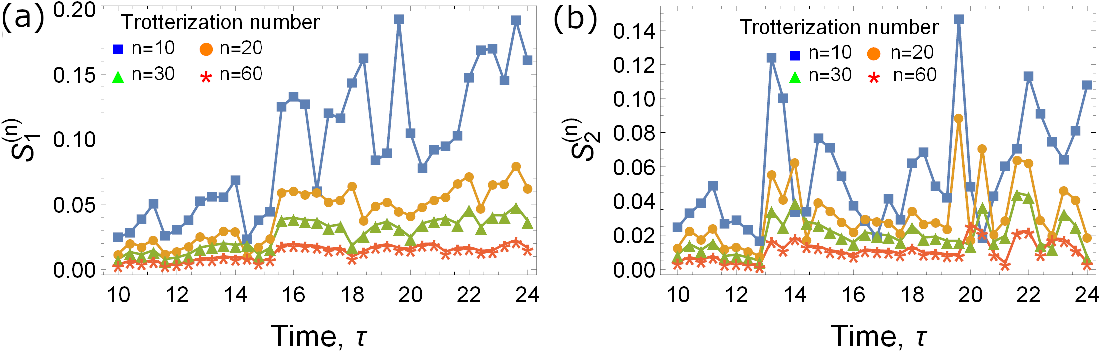}
\caption{The $\tau$-dependence of the characteristics $S^{(n)}_1$
(a) and $S^{(n)}_2$ (b). Here $N=6$, $N^{(S)}=N^{(R)}=2$.}
\label{Fig:S1S2XXZ}
\end{figure*}

We consider the short chain of $N=6$ spins with
$N^{(S)}=N^{(R)}=N^{(TL)}=2$. We use  two nonzero Larmor
frequencies associated with two last nodes of the chain:
$\omega_{6}$ and $\omega_{5}$. Eq.~(\ref{omega2}) yields
$\omega_6=-\omega_5$, thus we have only one free Larmor frequency
$\omega_5$ which will be used as a control tool. The system
(\ref{constr_red}) consists of two complex equations and there are
only two independent scale factors $\lambda_{10}$ and
$\lambda_{20}$.  To solve (\ref{constr_red}), we need at least four
parameters $a_{5j}$, $j=1,\dots,4$, generated by the step-wise
$\tau$-dependence of $\omega_5$ as given in (\ref{LF}), therefore we
fix $K_\omega=4$ and $w=\{a_{51},\dots,a_{54}\}$ in
(\ref{constr_red_appr}). In addition,  $\Delta\tau=
{\tau_{\mathrm{reg}}/4}$. For optimization of $S_i^{(n)}$,
$i=1,\dots,4$, in Model~1, we take 1000 different solutions of
system (\ref{constr_red_appr}) $a^{(m)}_{5j}$ or $\tilde
a^{(m)}_{5j}$ (according to representation (\ref{ta})),
$j=1,\dots,4$, $m=1,\dots,1000$. We approximate the evolution
operator with the Trotterization formula (\ref{Vtr}) setting the
Trotterization number $n=10$, 20, 30, 60. Characteristics
$S^{(n)}_i$, $i=1,2$, are shown in Fig.~\ref{Fig:S1S2XXZ}. We see
that the Trotterization yields a good approximation for the
evolution over relatively short time intervals $\sim 24$ and error
of approximation decreases with an increase in Trotterization
number.

\begin{figure*}[!]
\centering
\includegraphics[width=\textwidth]{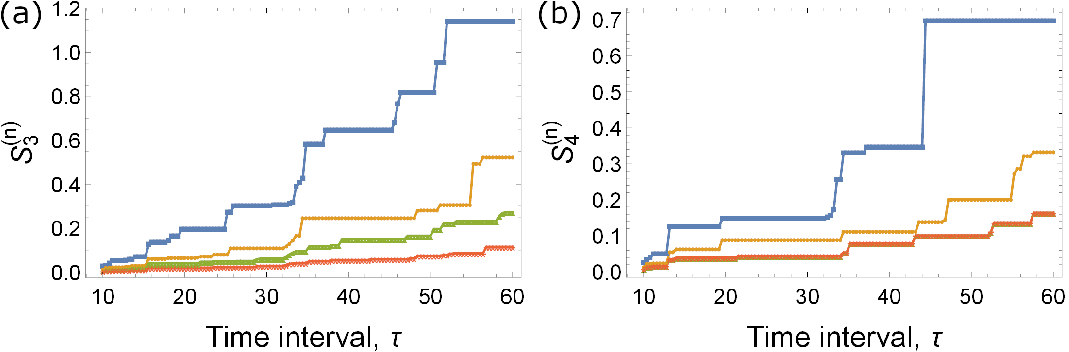}
\caption{The $\tau$-dependence of  the characteristics $S^{(n)}_3$
(a) and $S^{(n)}_4$ (b).  The four lines from bottom to up
correspond to the Trotterization number $n$  equal to $60$, $30$,
$20$, and $10$, respectively. Here $N=6$, $N^{(S)}=N^{(R)}=2$.}
\label{Fig:S3S4XXZ}
\end{figure*}

\begin{figure*}[!]
\centering
\includegraphics[width=0.5\textwidth]{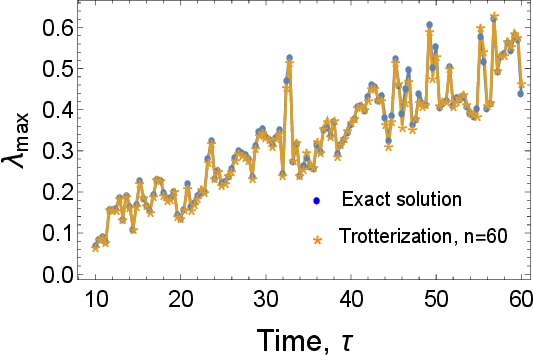}
\caption{The $\tau$-dependence of the maximal  scale factor
$\lambda_{\max}=\max(\lambda_{\{01\};\{00\}},\lambda_{\{10\},\{00\}})$.
Both the exact and Troterrized ($n=60$) evolutions are shown. Here
$N=6$, $N^{(S)}=N^{(R)}=2$. } \label{Fig:maxXXZ}
\end{figure*}

Approximation over longer time  intervals is worse, as shown by
$S_i$, $i=3,4$  in Fig.~\ref{Fig:S3S4XXZ}.
Comparison of the exact evolution of
$\lambda_{\max}=\max(\lambda_{\{01\};\{00\}},\lambda_{\{10\};\{00\}})$
with the approximated  evolution (with Trotterization number $n=60$)
is shown in Fig.~\ref{Fig:maxXXZ} for the optimized parameters
$\omega^{(n)}$. We see that the approximation is rather good over
the whole considered time interval $0\le{\tau_{\mathrm{reg}}}\le
60$.

\begin{figure*}[!]
\centering
\includegraphics[width=0.5\textwidth]{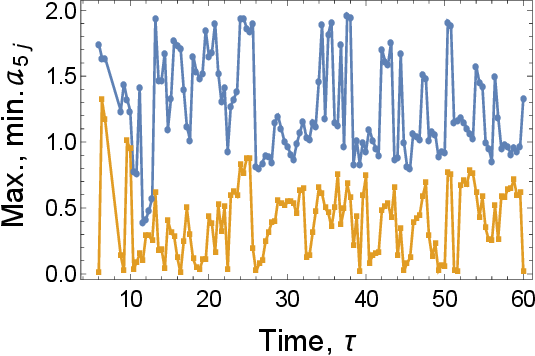}
\caption{The maximal (upper curve) and minimal (lower curve)  values
of $a_{5j}$, $j=1,\dots,4$ obtained in constructing $S_1^{(60)}$.
Here $N=6$, $N^{(S)}=N^{(R)}=2$. } \label{Fig:MMXXZ}
\end{figure*}

For completeness, we characterize the values  of the parameters
$a_{5j}$ used for  constructing $S_1^{(60)}$ (with the
Trotterization number $n=60$)  by representing the maximal and
minimal (by  absolute values) optimizing (minimizing  $S_1^{(n)}$)
parameters $a_{5j}$, $j=1,\dots,4$ ($\min_j\, |a_{5j}|$, $\max_j\,
|a_{5j}|$) as functions of time instant $\tau$ in
Fig.~\ref{Fig:MMXXZ}. By construction, these values are inside of
the interval $[0,2]$.

We just add  the parameter $\tilde \varepsilon$ into  the list of
the arguments of $S_i^{(n)}$ in formulae (\ref{epsilon})--(\ref{S5})
with $n=0$. Let us study the dependence of $S_i^{(0)}$,
$i=1,\dots,5$, on $\varepsilon$. We proceed as follows. First, we
set  $ \varepsilon=1/2$, $\Delta\tau^{(2)}= \Delta\tau^{(1)}$,
$\Delta\tau=2  \Delta\tau^{(1)}$ and  find the parameters
$\omega^{(0;m)}=\{a_{51}^{(m)},\dots,a_{54}^{(m)}\} $  as solutions
of  system (\ref{constr_red_appr}). It is remarkable that having
$\Omega_j$ constructed with $\varepsilon =1/2$ ($\Delta\tau^{(2)}=
\Delta\tau^{(1)}$), we can deform the obtained result to various
$\varepsilon$ with appropriate shrinking of $\tau$ without
additional  simulating spin dynamics. For a further analysis, it is
convenient to introduce the parameter $\tilde\varepsilon$ instead of
$\varepsilon$ given in  (\ref{var0}) and
$\tilde\varepsilon$-dependent subinterval
$\Delta\tau^{(2)}(\tilde\varepsilon)$ by the formulae
\begin{eqnarray}\label{te}
\Delta\tau^{(2)}(\tilde\varepsilon)  =\Delta\tau_1 \tilde\varepsilon \;\;\Rightarrow \;\;
\tilde \varepsilon = \frac{\Delta\tau^{(2)}(\tilde\varepsilon) }{\Delta\tau_1}.
\end{eqnarray}
Then, we have
\begin{eqnarray}\nonumber
\Omega_j \Delta \tau^{(1)} = \frac{\Omega_j}{\tilde\varepsilon} \Delta\tau^{(2)}(\tilde\varepsilon).
\end{eqnarray}
Therefore, we may use  the replacement  $\displaystyle a_{ij} \to
\frac{a_{ij}}{\tilde\varepsilon}$  in the Hamiltonian (\ref{HHj}),
so that condition (\ref{avar}) is satisfied and this condition
provides  the structure (\ref{HHj2}) for the Hamiltonian. Because of
(\ref{te}), the interval $\Delta\tau$ becomes also
$\tilde\varepsilon$-dependent,
\begin{eqnarray}
\nonumber
\Delta \tau(\tilde\varepsilon) = \Delta\tau^{(1)} +\Delta\tau^{(2)}(\tilde\varepsilon).
\end{eqnarray}
It can be simply verified that $ \tilde \varepsilon =
\varepsilon/(1-\varepsilon)\stackrel{\varepsilon\ll 1}{ \approx}
\varepsilon . $ In this way, we can vary $\tilde \varepsilon$ by
scaling $\Omega_j$  and thus scale $a_{ij}$. In studying the
functions $S_i^{(0)}$, $i=1,\dots,5$, below we use
$\tilde\varepsilon$ instead of $\varepsilon$.

\begin{figure*}[!]
\centering
\includegraphics[width=\textwidth]{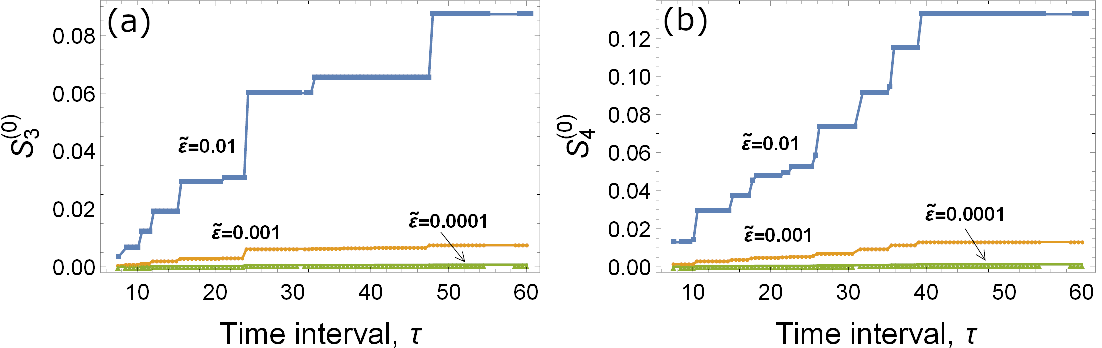}
\caption{The $\tau$-dependence of  the characteristics $S^{(0)}_3$
(a)  and $S^{(0)}_4$ (b) for different  $\tilde \varepsilon =0.01,\;
0.001,\; 0.0001$. Here $N=6$ and $N^{(S)}=N^{(R)}=2$. }
\label{Fig:2S3S4XXZ}
\end{figure*}

The graphs of $S^{(0)}_{i}$, $i=1,\dots,4$, are quite similar  to the
graphs of $S^{(n)}_{i}$, $i=1,\dots,4$, $n>0$. For instance, the
graphs of $S^{(0)}_{i}$, $i=3,4$ as functions of  $\tau$ for
different  $\tilde \varepsilon =0.01,\; 0.001,\; 0.0001$ are shown
in Fig.~\ref{Fig:2S3S4XXZ}.

Notice that since the parameters  $a_{5j}$ are represented  in the
form~(\ref{ta}), they are  bounded by the condition $|a_{5j}|\in
[0,2]$. Of course, the scaled parameters
$\displaystyle\frac{a_{5j}}{\tilde\varepsilon}$ do not obey this
constraint.

\paragraph{Entanglement transfer in state restoring}.

\begin{figure*}[!]
\includegraphics[width=\textwidth]{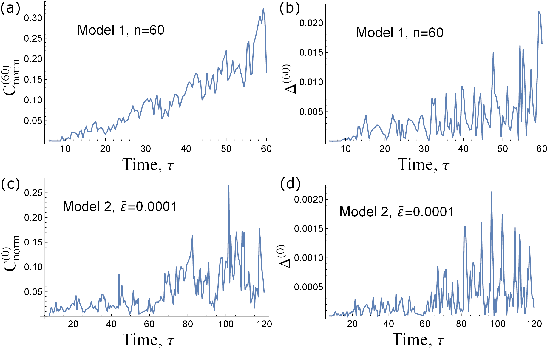}
\caption{(a) Model~1 with $n=60$, ratio $C^{(60)}_{\mathrm{norm}}$ (b)
Model~1 with $n=60$, the relative discrepancy $\Delta^{(60)}$ (c)
Model~2 with  $\tilde \varepsilon = 0.0001$, the ratio
$C^{(0)}_{\mathrm{norm}}$   (d)  Model~2 with $\tilde \varepsilon =
0.0001$, the relative  discrepancy $\Delta^{(0)}$.}
\label{Fig:conc}
\end{figure*}

The 2-qubit (0,1)-excitation density matrix is  a $4\times 4$ matrix
in the form
\begin{eqnarray}\nonumber
\rho=\left(\begin{array}{cccc}
\rho_{00}&\rho_{01}&\rho_{02}&0\cr
\rho_{10}&\rho_{11}&\rho_{12}&0\cr
\rho_{20}&\rho_{21}&\rho_{22}&0\cr
0&0&0&0
\end{array}\right).
\end{eqnarray}
It was shown in \cite{TZarxiv} that the bi-particle entanglement
measured by the Wootters criterion \cite{HW,Wootters} in terms of
concurrence can be simply expressed in terms of single element of
the 2-qubit  density matrix in the case of (0,1)-excitation (see
also \cite{AQJ,FYu}), $ C= 2 |\rho_{12}|.$ In the case of state
restoring, when $\rho^{(R)}_{12}= \lambda_{12} \rho^{(S)}_{12}$, the
ratio $C^{(R)}/C^{(S)} = 2|\lambda_{12}|$ does not
depend on the sender's initial state   \cite{TZarxiv}. However, since we deal with
approximation  models, Model~1 and Model~2,  the above statement
holds only for them
\begin{eqnarray}\nonumber
C_{\mathrm{norm}}^{(n)}(\tau)=
\frac{C(\rho^{(R;n)}(\tau))}{C(\rho^{(S)}(0)) }=2|\lambda_{12}|,
\quad
C_{\mathrm{norm}}(\tau)=\frac{C(\rho^{(R)}(\tau))}{C(\rho^{(S)}(0))}
=f(\rho^{(S)}(0)).
\end{eqnarray}
Here $\rho^{(R;n)}$ means the density matrix  of the 2-qubit
receiver state calculated for  Model~1 ($n>0$) or Model~2 ($n=0$).
The $\tau$-evolution of $C^{(n)}_{\mathrm{norm}}$ is shown in
Fig.~\ref{Fig:conc}a (Model~1, $n=60$) and in Fig.~\ref{Fig:conc}c
(Model~2, $\tilde\varepsilon=0.0001$) for the EPR initial sender's
state
\begin{eqnarray}\nonumber
\rho^{(S)}(0)=\frac{1}{2} (|01\rangle +|10\rangle)  (\langle
01|+\langle 10|),\quad C(\rho^{(S)}(0))=1.
\end{eqnarray}
We also introduce the relative discrepancy between  the concurrences
calculated for  Model~1 (Model~2) and for exact solution,
\begin{eqnarray}\nonumber
\Delta^{(n)}(\tau)=\frac{\left|C(\rho^{(R;n)}(\tau))- C(\rho^{(R)}(\tau))\right|}{C(\rho^{(S)}(0))},
\end{eqnarray}
whose $\tau$-dependence is shown in Fig.~\ref{Fig:conc}b   (Model~1,
$n=60$)  and Fig.~\ref{Fig:conc}d (Model~2,
$\tilde\varepsilon=0.0001$) for the same EPR state.
Figures~\ref{Fig:conc}a and \ref{Fig:conc}c  demonstrate that
evolution of concurrence  in Model~2  is slower than in Model~1 and
exhibits high-amplitude oscillations.

\subsubsection{Two- and three-qubit state restoring in long chains ($N\le
20$).} To demonstrate the applicability of the  state-restoring
protocol to long  communication lines, we consider the
$N$-dependence of the  function $S^{(0)}_5$ in the state-restoring
of two- and three-qubit states transferred using communication lines
with $N\le 20$. In this section we use only Model~2 and take $\tilde
\varepsilon=0.0001$, i.e., $\Delta\tau \approx \Delta\tau^{(1)}$.

\begin{figure*}[!]
\centering
\includegraphics[width=\textwidth]{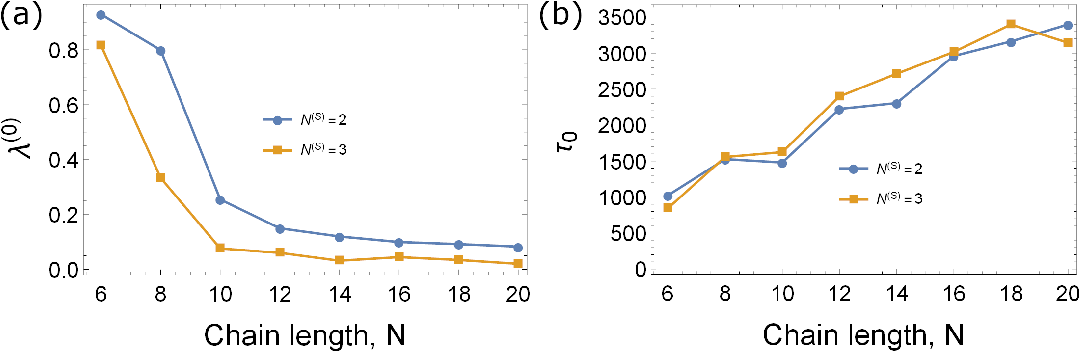}
\caption{$N^{(S)}=N^{(R)}=2$ and 3; the scale factor
$\lambda^{(0)}(N)=S_5^{(0)}(\tau_0,N)$, $0\le \tau_0(N) \le 200 N$,
(a) and the corresponding  time instant $\tau_0(N)$ (b)  as
functions of the chain length $N$; optimization is over 1000
solutions of Eq.~(\ref{constr_red_appr}) ($n=0$) for $N^{(S)}=2$ and
over 2000 solutions for $N^{(S)}=3$. } \label{Fig:CTXXZ}
\label{Fig:CT2XXZ}
\end{figure*}

\paragraph{Two-qubit sender, $N^{(S)}=N^{(R)}=2$.}
We plot  $\lambda^{(0)}$, given in (\ref{lamn2}), with fixed
$K_\omega=5$,   as a function of $N$ in Fig.~\ref{Fig:CTXXZ}. For
optimization  $S_5^{(0)}(T,N)$, we take $T=200 N$ and 1000 different
solutions $\omega^{(0;m)}$ of the system~(\ref{constr_red_appr}) at
$n=0$.
These graphs demonstrate generic decrease of  $\lambda^{(0)}(N)$ and
increase of $\tau_0(N)$ with $N$. The fact that points of these
graphs do not belong to smooth lines is explained, basically, by
difficulties in the global optimization  of $S^{(0)}_5$ according to
its definition~(\ref{S5}).

\paragraph{Three-qubit sender, $N^{(S)}=N^{(R)}=3$.}
\label{Section:NnS3}

We take three nonzero Larmor frequencies,  $\omega_{7}$,
$\omega_{6}$ and $\omega_{5}$. Eq.~(\ref{omega2}) yields
$\omega_7=-\omega_6-\omega_5$, thus we have two free Larmor
frequency  $\omega_5$ and $\omega_6$  which can be used as a control
tool. The system (\ref{constr_red_appr}) consists of 6 complex
equations. We have 3 independent scale parameters $\lambda_{i,0}$,
$i=1,\;2,\;3$. To satisfy system (\ref{constr_red_appr}), we take
$K_\omega=7$, so that there are 14 free parameters $a_{5j}$,
$a_{6j}$,  $j=1,\dots,7$,  generated by  $\omega_5$ and $\omega_6$.
Eqs.~(\ref{constr_red_appr}) are considered as a system of equations
for these 14 parameters. For optimization in formulas for
$S_5^{(0)}(T,N)$, we take $T=200 N$ and 2000 different solutions of
system (\ref{constr_red_appr}) ($n=0$) $a^{(m)}_{5j}$,
$a^{(m)}_{6j}$ or $\tilde a^{(m)}_{5j}$, $\tilde a^{(m)}_{6j}$
(according to representation (\ref{ta})), $j=1,\dots,7$,
$m=1,\dots,2000$. Finally, we plot  $\lambda^{(0)}$ (\ref{lamn2})
and appropriate $\tau_0$ as  functions of $N$ in Fig.
\ref{Fig:CT2XXZ}. We see that the line $\lambda^{(0)}$ for
$N^{(S)}=3$ is below the line for $N^{(S)}=2$ in Fig.
\ref{Fig:CT2XXZ}a, unlike the lines for $\tau_0$ in Fig.
\ref{Fig:CT2XXZ}b.

\section{Perfect transfer of  0-order coherence matrix along zigzag- and~rectangular (multichannel) communication lines }
\label{Section:0cm} 
In the recent paper \cite{BFLPZ_2022} the
perfect transfer of zero-order coherence matrix (PTZ) including up
to 1-excitation block governed by the XX-Hamiltonian was studied.
That was a simplest nontrivial case, but that PTZ doesn't allow to
include higher-order ($\pm 1$-order) coherence matrices in the
process. To make it possible we  involve blocks of higher excitation
numbers into the density matrix of the initial state. Namely, we
consider the initial state of tripartite system including sender
($S$), transmission line ($TL$) and receiver ($R$) in form
(\ref{rho0}), (\ref{TLR}) and assume that the dynamics is governed
by XXZ-Hamiltonian (\ref{HH0}), (\ref{XX}) with $\Omega=0$. In the
case when initial sender state $\rho^{(S)}$ includes only 0-order
coherence matrix so that it is a block-diagonal initial sender
state, $ \rho^{(S)} = \bigoplus_{i=0}^K s^{(i)}, $ where $s^{(i)}$
is the $i$-excitation block. Thus, the density matrix of the whole
$N$-qubit system has the similar block-diagonal form: $ \rho =
\bigoplus_{i=0}^K \sigma^{(i)} , $ where $\sigma^{(i)}$ for
$N$-qubit system is a $D^{(N)}_i\times D^{(N)}_i$ matrix,
$D^{(N)}_i=\left({N}\atop{i}\right)$. Here $K$ is the maximal number of
excitations which is restricted by the requirement $K\le N^{(S)}$.
Similarly, for the  receiver density matrix ($N^{(R)}=N^{(S)}$) we
have $ \rho^{(R)} = {\mbox{Tr}}_{S,TL}\rho(t) = \bigoplus_{i=0}^K
r^{(i)}(t). $

\subsection{PTZ using  whole state space of sender}
\label{Section:PTZfull} In this case, $K=N^{(S)}$. It was shown in
\cite{BFLPZ_2022}, that PTZ in the full state space of sender
involves the unitary transformation $U^{\mathrm{ex}}$ (where ``ex''
means ``exchange'') that exchanges two single-element blocks of the
receiver's state space, which are  0- and $N^{(S)}$-excitation
blocks, and does not conserve the excitation number, i.e.,
\begin{eqnarray}\nonumber
[U^{\mathrm{ex}},I_z]\neq 0.
\end{eqnarray}
In this case, the process  of the PTZ is most simple and includes
two steps.
\begin{enumerate}
\item
Register the state $\rho^{(R)}$ at some optimal time instant $\tau_{\mathrm{reg}}$ and solve the system
\begin{eqnarray} \label{syst1}
&&
r^{(i)}(\tau_{\mathrm{reg}}) = s^{(i)},\quad i=1,2,\dots, N^{(S)}-1,\\
\nonumber
&&
r^{(N^{(S)})}(\tau_{\mathrm{reg}}) = s^{(0)}
\end{eqnarray}
for the elements of blocks $s^{(i)}$, $i=0,1,\dots,N^{(S)}-1$. Then, due to the normalization, the following equality holds:
$
r^{(0)}(\tau_{\mathrm{reg}}) = s^{(N^{(S)})}.
$
\item
Exchange the single-element blocks,
$
r^{(0)}(\tau_{\mathrm{reg}}) \leftrightarrow r^{(N^{(S)})}(\tau_{\mathrm{reg}})
$,
using the transformation $U^{\mathrm{ex}}$, which in this case is given by
\begin{eqnarray}\nonumber
U^{ex}=\left(
\begin{array}{c|c|c}
0&0_{1,2^{N^{(S)}}-2}& 1\cr
\hline
0_{2^{N^{(S)}}-2,1}& \mathbb I_{2^{N^{(S)}}-2,2^{N^{(S)}}-2}&0_{2^{N^{(S)}}-2,1}\cr
\hline
1&0_{1,2^{N^{(S)}}-2} &0
\end{array}
\right),
\end{eqnarray}
where $0_{i,j}$ and $\mathbb I_{ij}$ are, respectively, $i\times j$ zero and unit matrices.
\end{enumerate}
As the result, we obtain $ \rho^{(R)}(\tau_{\mathrm{reg}}) =
\rho^{(S)}(0). $ The time instant for state registration can  be
selected by some additional requirement discussed below. It is
important that PTZ in this case can be organized without unitary
transformation of the extended receiver.

\subsection{PTZ using cut state space of sender}
\label{Section:PTZcut} Now, let $K<N^{(S)}$, then only the
0-excitation  block is a single-element block. Therefore, we have to
create second single-element diagonal block  \cite{BFLPZ_2022} which
will be exchanged with the 0-excitation block by the appropriate
transformation $U^{(ex)}$. In principle, any diagonal element in any
of two blocks $s^{(1)}$ or  $s^{(2)}$ can be singled out by putting
zeros for all nondiagonal elements of appropriate row and column.
For instance, let the diagonal element of $s^{(2)}$ corresponding to
the 1st and 2nd excited qubits be such an element
\begin{eqnarray}\label{s2}
s^{(2)} = \left(
\begin{array}{c|c}
s^{(2)}_{11} & 0_{1, D^{(N^{(S)})}_2{-1}}\cr
\hline
0_{ D^{(N^{(S)})}_2{-1},1}&\tilde s^{(2)}
\end{array}
\right),\quad D^{(N^{(S)})}_2=\left({N^{(S)}}\atop{2}\right),
\end{eqnarray}
where $0_{ij}$ is the $i\times j$ zero matrix and $\tilde s^{(2)}$ is a $(D^{(N^{(S)})}_2-1) \times (D^{(N^{(S)})}_2-1)$ matrix.

Evolution under the operator $V(\tau)$ in the block-diagonal   form
mixes all elements in each $i$-excitation block of the density
matrix. Therefore, in general, the block $r^{(2)}$ in the density
matrix of the receiver $\rho^{(R)}$  has all nonzero elements
\begin{eqnarray}
\nonumber
r^{(2)} = \left(
\begin{array}{c|c}
r^{(2)}_{1,1} &\bar{ r}^{(2)}\cr
\hline
\Big(\bar{ r}^{(2)}\Big)^\dagger&\tilde r^{(2)}
\end{array}
\right) ,
\end{eqnarray}
where $\bar{ r}^{(2)}$ is a row of $D^{(N^{(S)})}_2{-1}$ elements
$\bar{ r}^{(2)}_j \equiv  r^{(2)}_{1{,}j{+1}},\;\;j=1,\dots,
D^{(N^{(S)})}_2{-1}$, $\tilde r^{(2)}$ is a $(D^{(N^{(S)})}_2-1)
\times (D^{(N^{(S)})}_2-1)$ matrix,  and we use the fact that
$N^{(R)}=N^{(S)}$. To single out the element $r^{(2)}_{11}$ as a
one-element block on the diagonal  we have to introduce the unitary
transformation $U(\varphi)$ of the extended receiver $ER$ (the
receiver with several neighboring nodes of the transmission line) at
time instant $\tau_{\mathrm{reg}}$. Thus, the free  parameters
$\varphi$ of the unitary transformation $U$ appear in the elements
of $\rho^{(R)}$. Then, we can  solve the system
\begin{eqnarray}\label{req1}
&&
\bar{r}^{(2)}=0,\\\label{req2}
&&
\tilde r^{(2)} = \tilde s^{(2)},\quad
r^{(1)}=s^{(1)},\quad
r^{(2)}_{11}=s^{(0)}.
\end{eqnarray}
In addition, due to the trace-normalization
\begin{eqnarray}\nonumber
s^{(0)}+s^{(2)}_{11}+\sum_j s^{(1)}_{jj} + \sum_j \tilde s^{(2)}_{jj}  =
r^{(0)}+r^{(2)}_{11}+\sum_j r^{(1)}_{jj} + \sum_j \tilde r^{(2)}_{jj}  =
1,
\end{eqnarray}
we have $ r^{(0)}=s^{(2)}_{11}. $ To solve system (\ref{req1}) and
(\ref{req2}), we notice that  subsystem (\ref{req2}) is a linear
system in the elements of $s^{(1)}$ and  $\tilde s^{(2)}$,
$s^{(0)}$. Solving this subsystem and substituting expressions for
$s^{(1)}$, $\tilde s^{(2)}$ and $s^{(0)}$  into  system
(\ref{req1}), we obtain a system of nonlinear equations for the
parameters  $\varphi$ of the unitary transformation of the extended
receiver (don't mix with $U^{\mathrm{ex}}$). Finally  we exchange $
r^{(0)} \leftrightarrow r^{(2)}_{11} $ using $U^{\mathrm{ex}}$.
Then, the state of the receiver at the selected time instant
$\tau_{\mathrm{reg}}$ is equal to the initial sender's state: $
\rho^{(R)}(\tau_{\mathrm{reg}})=\rho^{(S)}(0). $

{\bf Remark.} Solution of system  (\ref{req1}) and (\ref{req2}) is
not unique due to the presence of extra parameters $\varphi$ in the
unitary transformation. A  particular solution can be found
relatively simply  solving Eq.~(\ref{req1}) for $\varphi$  (make it
identity for all elements of $\rho^{(S)}$) and then solving linear
system (\ref{req2}) with constant coefficients for the elements of
$\rho^{(S)}$. In fact, $r^{(2)}$ can be represented as
\cite{TZarxiv}
\begin{eqnarray}\label{r22}
r^{(2)} = {\cal{W}} s^{(2)} {\cal{W}}^\dagger,
\end{eqnarray}
where ${\cal{W}}$ is some  matrix. Therefore, Eq.~(\ref{req1}) can be given the following
form
\begin{eqnarray}\nonumber
\bar{r}^{(2)}_{j}\equiv  r^{(2)}_{1j} =\sum_{k,n} {\cal{W}}_{1k} {\cal{W}}^*_{jn}s^{(2)}_{kn}.
\end{eqnarray}
Then, (\ref{req1}) holds if
\begin{eqnarray}\label{calW}
 {\cal{W}}_{1k}=0,\quad k=1,\dots, D^{(N^{(S)})}_2.
\end{eqnarray}

\subsection{Unitary transformation}
Let $W$ be the superposition of the evolution $V(\tau)=e^{-i H \tau
/D_{12} }$  and unitary transformation of the extended receiver
$U(\varphi)$,
\begin{eqnarray}\nonumber
W(\varphi,\tau)=(\mathbb I\otimes U(\varphi)) V(\tau),
\end{eqnarray}
where $\mathbb I$ is the  unit matrix in the space of the whole
system without the extended receiver.  The unitary operators $U$ and
$W$ have the block-diagonal form
\begin{eqnarray}\nonumber
U&=&\bigoplus_k U^{(k)}(\varphi^{(k)}),\quad U^{(0)}=1,\\\nonumber
W&=&\bigoplus_k W^{(k)}(\Phi^{(k)}),\quad W^{(0)}=1,\quad \Phi^{(k)}=\{\varphi^{(1)},\dots,\varphi^{(k)}\},
\end{eqnarray}
where $U^{(i)}$ and $W^{(i)}$ are the blocks corresponding to  the
subspace of $i$ excitations. The equality $U^{(0)}=1$ is provided by
the freedom in the energy of the ground state.

We introduce the multiindexes (\ref{MI}). Then, we can write
\begin{eqnarray}\nonumber
\rho^{(R)}=\sum_{N_S,N_{TL}} W_{N_S N_{TL} N_R;I_S 0_{TL} 0_R} \rho^{(S)}_{I_SJ_S}
W^\dagger_{J_S, 0_{TL} 0_R;N_SN_{TL} M_R}.
\end{eqnarray}
The number of excitation in the block-diagonal  structure of the
density matrix varies from 0 to $N^{(S)}$ and thus depends on the
dimension of the sender.

\subsubsection{$N^{(S)}=3$,  (0,1,2,3)-excitation  state space.} In the case of 3-qubit sender, we can effectively use up
to  3-excitation states, therefore  we have only 4 blocks
\begin{eqnarray}\label{R3}
r^{(3)}= W^{(3)}_{0_S 0_{TL}  1_R; 1_S 0_{TL} 0_R}(\Phi^{(3)})
s^{(3)}_{1_S 1_S}  (W^{(3)})^\dagger_{1_S 0_{TL} 0_{R};0_S 0_{TL}
1_R}(\Phi^{(3)}),
\end{eqnarray}
\begin{eqnarray}\label{R2}
&&r^{(2)}_{N_RM_R}= \sum_{|N_S|+|N_{TL}|=1} W^{(3)}_{N_S N_{TL}
N_R; 1_S 0_{TL} 0_R}(\Phi^{(3)}) s^{(3)}_{1_S 1_S}
(W^{(3)})^\dagger_{1_S 0_{TL} 0_{R};N_S N_{TL}  M_R}(\Phi^{(3)}) \\
\nonumber && +\sum_{{I_S,J_S}\atop{|I_S|=|J_S|=2}} W^{(2)}_{0_S
0_{TL}  N_R; I_S 0_{TL} 0_R}(\Phi^{(2)}) s^{(2)}_{I_S J_S}
(W^{(2)})^\dagger_{J_S 0_{TL} 0_{R};0_S 0_{TL}  M_R}(\Phi^{(2)}),
\\\nonumber &&\hspace{3cm} |N_R|=|M_R|=2,
\end{eqnarray}
\begin{eqnarray}\label{R1}
&& r^{(1)}_{N_RM_R}= \sum_{|N_S|+|N_{TL}|=2} W^{(3)}_{N_S N_{TL}
N_R; 1_S 0_{TL} 0_R}(\Phi^{(3)})  s^{(3)}_{1_S 1_S}
(W^{(3)})^\dagger_{1_S 0_{TL} 0_{R};N_S N_{TL}  M_R}(\Phi^{(3)})
\\\nonumber && +\sum_{|N_S|+|N_{TL}|=1} \sum_{|I_S|=|J_{S}|=2}
W^{(2)}_{N_S N_{TL}  N_R; I_S 0_{TL} 0_R}(\Phi^{(2)}) s^{(2)}_{I_S
J_S} (W^{(2)})^\dagger_{J_S 0_{TL} 0_{R};N_S N_{TL}
M_R}(\Phi^{(2)})\\\nonumber &&+\sum_{|I_S|=|J_{S}|=1} W^{(1)}_{0_S
0_{TL}  N_R; I_S 0_{TL} 0_R}(\Phi^{(1)}) s^{(1)}_{I_S J_S}
(W^{(1)})^\dagger_{J_S 0_{TL} 0_{R};0_S 0_{TL}
M_R}(\Phi^{(1)}),\\\nonumber &&\hspace{3cm}|N_R|=|M_R|=1,
\end{eqnarray}
\begin{eqnarray}
\nonumber
r^{(0)} =1-{\mbox{Tr}}(r^{(1)}) -{\mbox{Tr}}(r^{(2)})-r^{(3)}.
\end{eqnarray}

In the case of three-qubit sender, the system (\ref{syst1})  reduces
to the following one
\begin{eqnarray} \label{3syst1}
&&
r^{(i)}(\tau_{\mathrm{reg}}) = s^{(i)},\quad i=1, 2,\quad
r^{(3)}(\tau_{\mathrm{reg}}) = s^{(0)}.
\end{eqnarray}
Solving this system for the elements of $s^{(i)}$ we obtain the perfectly transferable 0-order coherence matrix.

\subsubsection{$N^{(S)}\ge 3$,  (0,1,2)-excitation  state space.} \label{Section:par_full} Now we consider initial
state with no more then 2 excitations.  In this case
$r^{(3)}=s^{(3)}=0$, eq.(\ref{R3}) disappears, while equations
(\ref{R2}) and (\ref{R1}) get the following form
\begin{eqnarray}\label{R2c}
&&r^{(2)}_{N_RM_R}= \sum_{|I_S|=|J_S|=2}W^{(2)}_{0_S 0_{TL}  N_R;
I_S 0_{TL}  0_R}(\Phi^{(2)}) s^{(2)}_{I_S J_S}
(W^{(2)})^\dagger_{J_S 0_{TL} 0_{R};0_S 0_{TL}  M_R}(\Phi^{(2)}),
\\\nonumber &&\hspace{3cm} |N_R|=|M_R|=2,
\end{eqnarray}
\begin{eqnarray}
\nonumber && r^{(1)}_{N_RM_R}= \sum_{|N_S|+|N_{TL}|=1}
\sum_{|I_S|+|J_{S}|=2}  W^{(2)}_{N_S N_{TL}  N_R; I_S 0_{TL}
0_R}(\Phi^{(2)})  s^{(2)}_{I_S J_S} (W^{(2)})^\dagger_{J_S 0_{TL}
0_{R};N_S N_{TL}  M_R}(\Phi^{(2)})\\\nonumber
&&+\sum_{|I_S|+|J_{S}|=1} W^{(1)}_{0_S 0_{TL}  N_R; I_S 0_{TL} 0_R}
s^{(1)}_{I_S J_S}(\Phi^{(1)}) (W^{(1)})^\dagger_{J_S 0_{TL}
0_{R};0_S 0_{TL}  M_R}(\Phi^{(1)}),\\\nonumber &&\hspace{3cm}
|N_R|=|M_R|=1.
\end{eqnarray}
Note that Eq.~(\ref{R2c}) is equivalent to Eq.~(\ref{r22})  up to
the change of notation $W^{(2)}\to {\cal{W}}$. To arrange the
structure (\ref{s2}) we require
\begin{eqnarray}\nonumber
W^{(2)}_{0_S 0_{TL} \{110\dots 0\}; I_S 0_{TL} 0_R}(\Phi^{(2)}) =0,
\end{eqnarray}
which is equivalent to (\ref{calW}).

\subsection{Zigzag spin chain, $N^{(S)}=3$}\label{Section:dm}

\begin{figure*}[!]
\epsfig{file=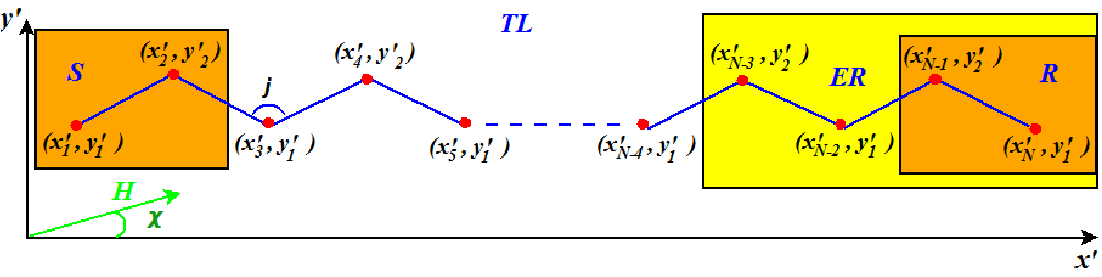,width=\textwidth,angle=0}
\caption{Zigzag spin 1/2 chain.}
\label{Fig:Z}
\label{Fig:CL}
\end{figure*}
We use the $XXZ$ Hamiltonian (\ref{XX}) with $D_{ij}$ taking into account the orientation of the external magnetic field,
\begin{eqnarray}\label{XXZ}
&&
H_{XXZ}=\sum_{j>i}
\frac{\gamma^2 }{2 r^3_{ij}} (3 \cos^2\varphi_{ij}-1)(X_i X_j + Y_iY_j - 2 Z_i Z_j ).
\end{eqnarray}
Let $j$th spin have coordinates $(x_j,y_j)$,  the magnetic field
${\mathrm H}$ is directed at the angle $\chi$ to the chain axis. In
the system $(x',y')$, the unit vector $\mathrm n$ has coordinates
${\mathrm n}=(\cos\chi,\sin\chi)$. Then,
\begin{eqnarray}
\label{coord2}
&&
{\mathrm r}_{ij} = (x'_j-x'_i, y'_j-y'_i),\\
\nonumber && r_{ij} = \sqrt{(x'_j-x'_i)^2 +(y'_j-y'_i)^2 }, \quad
\cos\varphi_{ij}=\frac{(x'_j-x'_i) \cos \chi+(y'_j-y'_i) \sin
\chi}{r_{ij}}. \label{coord4}
\end{eqnarray}
We consider
\begin{eqnarray}
\nonumber x'_j = \Delta (j-1),\quad y'_j =\left\{\begin{array}{ll}
0,&j=1,3,5,\dots, \cr y_0,& j=2,4,6,\dots,
\end{array}
\right.
\end{eqnarray}
i.e., the $y'$-coordinate takes two values, either  $0$ or $y_0$. We
use $\chi$ and $y_0$ as two free parameters characterizing the chain
geometry, Fig.~\ref{Fig:Z}. The communication line includes  the
sender ($S$), transmission line ($TL$) and receiver ($R$), see
Fig.~\ref{Fig:CL}. In addition, the extended receiver ($ER$) serves
to correct the receiver's state via the unitary transformation
applied to it. In this section we use  the dimensionless time $ \tau
= {t \gamma^2 }/{\Delta^3}. $

We consider the 3-qubit sender and receiver in the  9-qubit
communication system  $(N=9)$ under the XXZ Hamiltonian (\ref{XXZ}).
We are aimed at optimizing the perfectly transferred 0-order
coherence matrix including  0-, 1- and 2-excitation blocks  (without
3-excitation block). We say that the matrix is optimized if its
minimal diagonal element $\delta_d$ in the 1- and 2-excitation
blocks (except the element which is subjected to the operation
$U^{(ex)}$ in n.~2 below) takes the maximum possible value. We solve
this problem in two steps.

\begin{figure*}[h!]
\includegraphics[width=\textwidth]{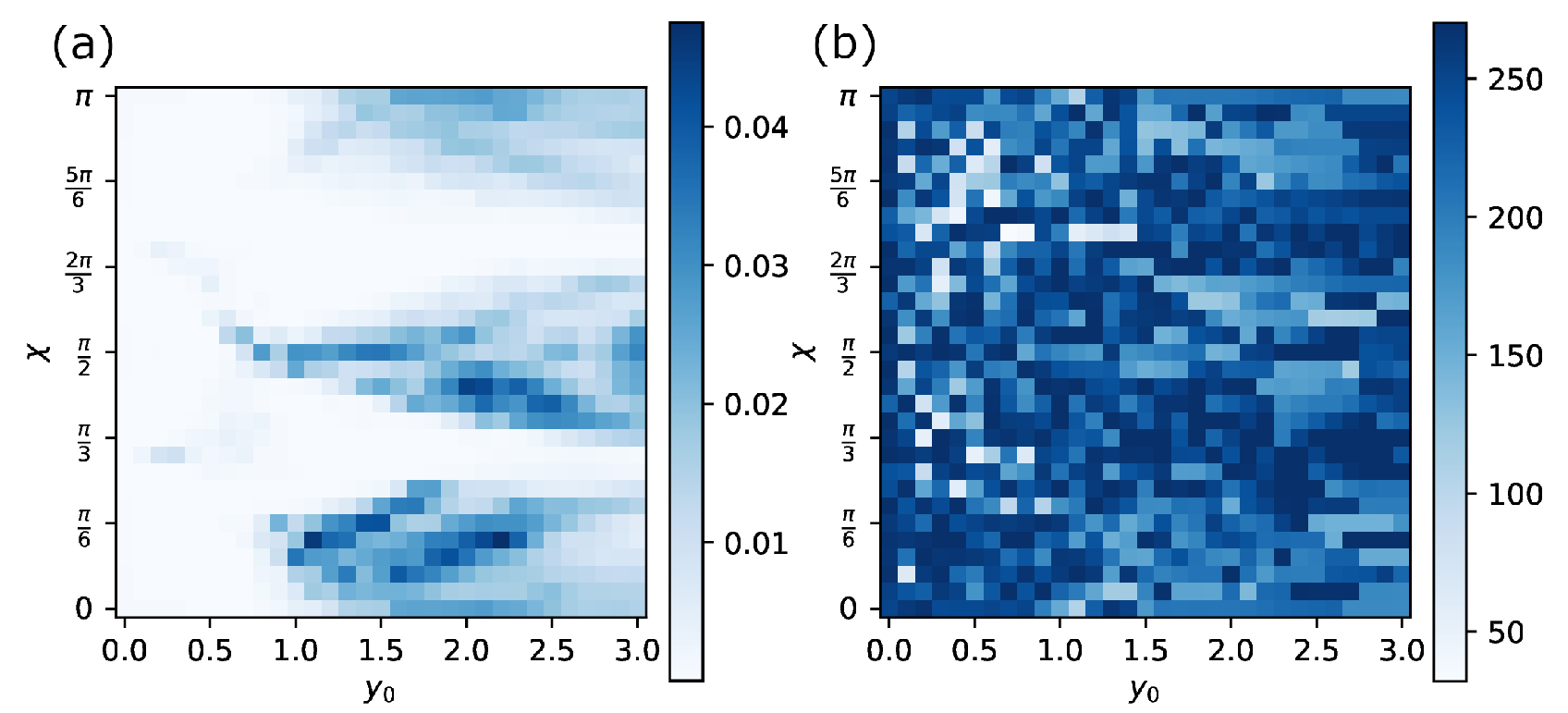}
\caption{(a) The parameter $\delta_d$ optimized over time $\tau$ and (b) corresponding optimal time instant $\tau_0$ as functions of  $y_0$ and $\chi$.
The region $\chi = \pi/2$,  $1 \lesssim y_0\lesssim 1.6$ can be considered to be optimal. In this case  $\delta_d\sim 0.35$.}
\label{fig10}
\end{figure*}

1. Let us consider the system with up to  3 excitations. In this
case, according to \cite{BFLPZ_2022}, elements with 0 and 3
excitations can be exchanged on the last step of restoring. For
fixed parameters  of the chain  $\chi$ and $y_0$, we find such
optimal time instant $\tau_0$ that maximizes  $\delta_d$. Thus, we
obtain the distributions of  $\delta_d$  and  $\tau_0$ on the plane
$(y_0,\chi)$ shown in Fig. \ref{fig10}.
The parameter $\delta_d$ has large values in the dark region on the
plane $(y_0,\chi)$ (see Fig.~\ref{fig10}a),  where $\chi = \pi/2$,
$1 \lesssim y_0\lesssim 1.6$. The same region  of large values of
$\delta_d$ remains for different lengths  of communication system
and for  cases of lower excitation space. In this case,
$\delta_d\sim 0.35$. We turn to this region.

2. Now we pass to the problem of optimal transfer of the 0-order
coherence matrix with up to 2-excitation blocks. In this case, to
provide structure (\ref{s2}) for the 2-excitation block, we have to
solve the system of equations  (\ref{req1}) and (\ref{req2}) for the
parameters of the unitary transformation of the extended receiver
(4-qubit subsystem in our case). After that we exchange positions of
two singled out elements of the  receiver and thus construct the
perfectly transferred 0-order coherence matrix including up to
2-excitation block with  $N^{(S)}=3$.

\begin{figure*}[!]
\includegraphics[width=\textwidth]{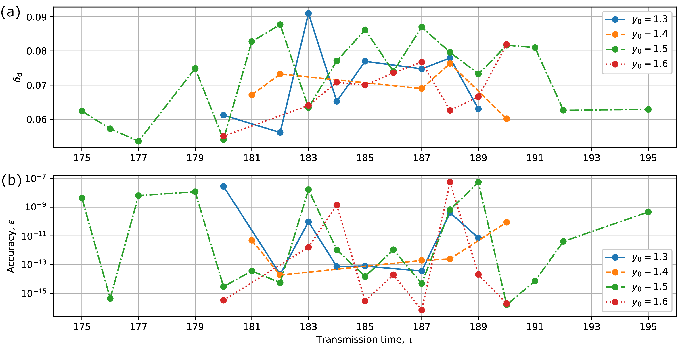} 
\caption{Time-dependence of (a) the parameter $\delta_d$  and (b) the accuracy of calculation of the transferred-to-zero elements for several values of the parameter $y_0$.}
\label{Fig:3}
\end{figure*}

Time-dependence of the parameter $\delta_d$, and  the accuracy of
transformation of the above pointed elements to zero are shown in
Fig.~\ref{Fig:3} for several values of the parameter $y_0$. We see
that all selected values of $y_0$ yield  much the same result, the
parameter $\delta_d$ takes the maximal value $\approx 0.9$ at
$\tau_{\mathrm{reg}} \approx 183$ for $y_0=1.3$. This is the
recommended time instant for the receiver's state registration.

\subsection{Rectangular (multichannel) spin chain}\label{Section:rc}

\begin{figure*}[!]
\includegraphics[width=0.5\textwidth]{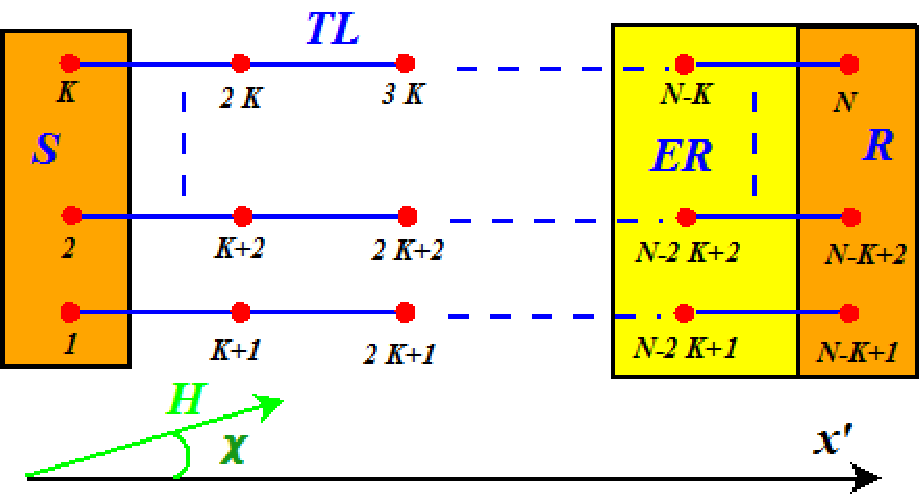}
\caption{Spin 1/2 lattice, $K$ is the number of channels.}
\label{Fig:rec}  \label{Fig:CL2}
\end{figure*}

Now we consider the $K$-channel lattice shown in Fig.~\ref{Fig:rec}.
The Hamiltonian is the same, Eq.~(\ref{XXZ}). The coordinates $x'_n$
and $y'_n$ of the $n$th spin  are following
\begin{eqnarray}\nonumber
x'_n=[(n-1)/K] \Delta_x,\quad y'_n = (n-1- K [(n-1)/K])\Delta_y,\quad \frac{N}{K} \in \mathbb Z,
\end{eqnarray}
where $[\cdot]$ means the integer part of a number. We introduce the
dimensionless time $ \tau = {t \gamma^2 }/{\Delta^3_x} $ and assume
that $\Delta_y$ and $\chi$ are free  parameters. Formulas
(\ref{coord2}) and  (\ref{coord4}) are applicable in this case as
well.

\begin{figure*}[!]
\includegraphics[width=\textwidth]{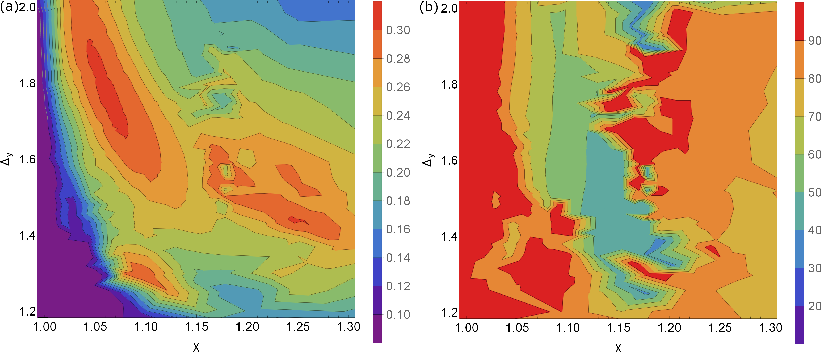}
\caption{(a) Parameter $\delta$ optimized over $\tau$ as a function  of
$\Delta_y$ and $\chi$; (b) the optimal time instant $\tau_0$ for the
state registration as a function of  $\Delta_y$ and $\chi$.}
\label{Fig:Rec}
\end{figure*}

We consider the propagation of the 3-qubit state along  the
three-channel communication line   ($K=3$ in Fig.\ref{Fig:rec}) with
10 spins in each channel (the 30-qubit system). Solving the system
(\ref{3syst1}), we find the $\tau$-dependence  for
$\rho^{(S)}(\tau)$. After that we find  $\tau_0$ that maximizes the
norm  $\delta(\tau,\Delta_y,\chi)$, similar to \cite{BFLPZ_2022}:
\begin{eqnarray}
\nonumber
\delta(\tau)=||\rho^{(R)}_\infty - \rho^{(R)}(\tau)||,
\end{eqnarray}
i.e., $ \label{deltamax} \delta(\tau_0)=\max_\tau \delta(\tau). $
Here $\rho^{(R)}_\infty={\mbox{diag}}(1,
\underbrace{0,\dots,0}_{2^K-1})$ is an asymptotic matrix
\cite{BFLPZ_2022}.  The contour graphs of
$\delta(\tau_0,\Delta_y,\chi)$ and of corresponding time instances
$\tau_0$ over the plane  $(\Delta_y,\chi)$ are  shown in Figs.
\ref{Fig:Rec}$a$ and \ref{Fig:Rec}$b$. The red region on the plane
$(\Delta_y,\chi)$ in Fig. \ref{Fig:Rec}$a$ ($\{1.04\lesssim \chi
\lesssim 1.08;1.65 \lesssim \Delta_y \lesssim 1.90 \}$) corresponds
to  $\delta\gtrsim 0.3$ at $\tau_0 \sim 60 \div 80$. The time
instant for state registration $\tau_{\mathrm{reg}}$ should  be
taken from that interval.

\section{Analytical approach}
\label{Section:analyt} The analytical
approach allows us to effectively study the long-time dynamics
which is especially important for the XXZ model which requires much
longer times for state transfer along a spin chain \cite{PRA2010}.
The analytical formulas were developed  in \cite{PRA2010} for the
homogeneous chain with remote end-nodes (in particular, the
homogeneous chain which we adopt here, see also \cite{FR,KF}) and
nearest-neighbor approximation, i.e.,
\begin{equation}
\nonumber
H_{XXZ}=\sum_{j=1}^{N-1}D(X_j X_{j+1} + Y_jY_{j+1} - 2 Z_j Z_{j+1}) .
\end{equation}
Due to the block-diagonal structure (\ref{bH})  of the Hamiltonian,
we can  consider the evolution in 0- and 1-excitation subspaces
separately.  The 1-excitation subspace is an $N\times N$ space, and
0-excitation subspace is a one-dimensional space.

\begin{figure*}[!]
\includegraphics[width=\textwidth]{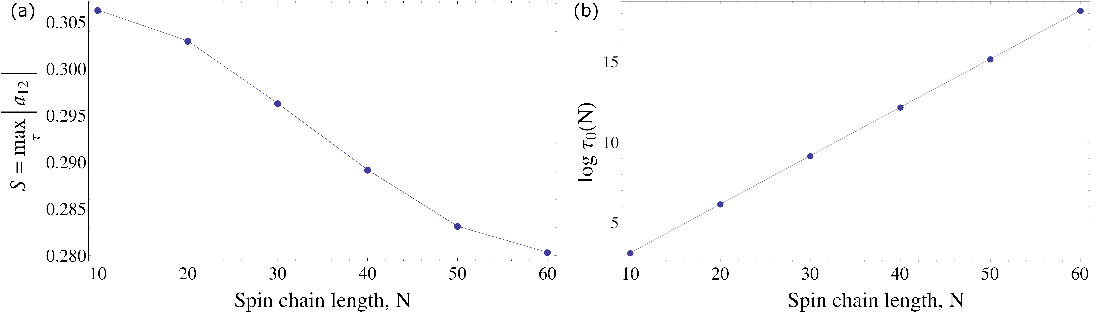}
\caption{$N$-dependence of the  parameter  (a) $S$  and (b) $\tau_{0}(S)$.}
\label{fig:maxampl}
\end{figure*}

We use the diagonalization of the  one-excitation  block $H^{(1)}$
proposed in \cite{PRA2010}. Elements of the eigenvector $u_j$ are
given by
\begin{equation}
\nonumber
u_{kj}= \left\{
\begin{array}{ll}\displaystyle
A_j\cos \frac{(N+1-2k)p_j}{2},& k=1,...N,\;\; j=1,\dots
N_1,\cr\displaystyle A_j\sin \frac{(N+1-2k)p_j}{2},&
k=1,...N,\;\;j=N_1+1,\dots N.
\end{array}\right.
\end{equation}
The appropriate eigenvalues are defined as follows: $
\label{eigenvalues} \lambda_j=D \cos p_j. $ Here the parameters
$p_j$ satisfy the following system
\begin{eqnarray}
\nonumber
 &&
2\cos{\frac{N-1}{2}p_j}+\cos{\frac{N+1}{2}p_j}=0,\quad j=1,\dots,
N_1,\quad N_1=N- \left[\frac{N}{2}\right],
\\\nonumber
&&  2\sin{\frac{N-1}{2}p_j}+ \sin{\frac{N+1}{2}p_j}  =0,\quad
j=N_1+1,\dots, N,
\end{eqnarray}
and the normalization constants  $A_j$ are given by
\begin{eqnarray}
\nonumber
A_j=\left\{
\begin{array}{ll}\displaystyle
\left( \frac{N-2}{2}+({1+\cos (N-1)p_j})+\frac{\sin (N-2)p_j}{2\sin
p_j} \right)^{-1/2},& j=1,\dots,N_1,\cr \displaystyle \left(
\frac{N-2}{2}+({1-\cos (N-1)p_j})-\frac{\sin (N-2)p_j}{2\sin p_j}
\right)^{-1/2} ,& j=N_1+1,\dots, N.
\end{array}\right.
\end{eqnarray}
We consider the 2-qubit sender and receiver  ($N^{(S)}=N^{(R)}=2$)
with the initial density matrix (\ref{rho0}) and   (\ref{TLR}),
where
\begin{eqnarray}\label{senderrho}
\rho^{(S)}(0)={\mbox{diag}}(a_{00},\rho^{(S;1)}),\quad
\rho^{(S;1)}=\begin{pmatrix}
a_{11} & a_{12}\\
a^*_{12} & a_{22}\\
\end{pmatrix}.
\end{eqnarray}
Here $a_{00}$ (scalar) and $\rho^{(S;1)}$  are blocks of the
sender's initial state including 0- and 1-excitation subspaces,
$a_{00}$ does not evolve. Evolution of the one-excitation block is
described by the operator $ U^{(1)}=\exp(-i{H^{(1)}} \tau /D) $,
where $\tau$ is the dimensionless time  (\ref{tau}), $D_{12}=D$. We
have
\begin{eqnarray}\label{fullrho}\nonumber
\rho(t)=\rho_{0,0}|0\rangle\langle 0|+\sum_{\stackrel{i, j=1}
{i',j'=1}}^N\sum_{m,m'=1}^2 u_{ij} u^*_{i'j'}
e^{-i(\lambda_j-\lambda_{j'}) \tau/ D} \rho^{(S)}_{m,m'}(0) u^*_{mj}
u_{m'j'} |i\rangle\langle i'|.
\end{eqnarray}
The receiver's density matrix (\ref{R}) reads
\begin{eqnarray}\nonumber
\rho^{(R)}(\tau_{\mathrm{reg}})=
\begin{pmatrix}\label{receiverrho0}
\rho_{0,0}^{(R)}(\tau_{\mathrm{reg}}) & 0 & 0\\
0 & \rho^{(R)}_{1,1}(\tau_{\mathrm{reg}}) & \rho^{(R)}_{1,2}(\tau_{\mathrm{reg}}) \\
0& (\rho^{(R)}_{1,2})^*(\tau_{\mathrm{reg}}) & \rho^{(R)}_{2,2}(\tau_{\mathrm{reg}})
\end{pmatrix}\nonumber\\
=\begin{pmatrix}\label{receiverrho}\nonumber
\rho_{0,0}+\sum_{j=1}^{N-2}{\rho_{j,j}(\tau_{\mathrm{reg}})} & 0 & 0\\
0 & \rho_{N-1,N-1}(\tau_{\mathrm{reg}}) & \rho_{N-1,N}(\tau_{\mathrm{reg}}) \\
0& \rho_{N,N-1}(\tau_{\mathrm{reg}}) & \rho_{N,N}(\tau_{\mathrm{reg}})
\end{pmatrix}.
\end{eqnarray}
Next, we fix the elements $a_{ij}$ in (\ref{senderrho})  by imposing
the following relation among the elements of
$\rho^{(R)}(\tau_{\mathrm{reg}})$, fixed at some time instant for
state registration $\tau_{\mathrm{reg}}$ (specified below), and the
elements $a_{ij}$:
\begin{eqnarray}
\nonumber
a_{12}= \rho^{(R)}_{12}(\tau_{\mathrm{reg}}),\quad
 a_{22} = \rho_{2,2}^{(R)}(\tau_{\mathrm{reg}}),\quad a_{00} = \rho_{11}^{(R)}(\tau_{\mathrm{reg}}).
\end{eqnarray}
As a characteristic of the remote state  restoring, we introduce
the function
\begin{eqnarray}
\nonumber
S(N)=\max_{0<\tau \lesssim 10^{3N/10}} |a_{12}(\tau)|.
\end{eqnarray}
We denote $\tau_{0}(N)$ the time instant $\tau$  corresponding to
$S(N)$, i.e., $S=|a_{12}(\tau_0(N))|$. $N$-dependence of  $S$ and
$\tau_{0}$  is shown in Fig.~\ref{fig:maxampl}. Thus, the
state-transfer time $\tau_{\mathrm{reg}}= \tau_0$ increases
exponentially with $N$.

\section{Conclusions}
\label{Section:Conclusions}

In this work we present several  aspects of evolution of spin chains
with XXZ-Hamiltonian. First, we consider the state-restoring
protocol with the time-dependent Larmor frequencies as a controlling
tool. To provide optimization of the remote state transfer, we
implement two approximating models. Model~1 uses Trotterization and
Model~2 uses short magnetic pulses of certain amplitude, similar to
that considered in \cite{FPZ_Archive2023} for the XX-Hamiltonian. We
use the quantities $S_i^{(n)}$ for $i=1,\dots,4$ to describe the
accuracy of the approximation and the quantity $S_5^{(n)}$ to
describe the quality of the state restoring, i.e., the absolute
values of $\lambda$-factors. This parameter  as a function of chain
length $N$ decreases with an increase of $N$ and, for fixed $N$,
reaches its maximal value at sufficiently long times, much longer
than for the XX-model. The confirmation of the long-time evolution
is presented in the theoretical consideration of the 0-order
coherence transfer, where exponential increase of the state-transfer
time with the chain length $N$ is found. We also study  the
two-qubit concurrence evolution  in the state-restoring process
and discrepancy in its calculation using an approximate model and
exact spin-chain evolution. The perfect transfer of the 0-order
coherence matrix including the 0-, 1-, 2- and 3-excitation blocks is
also considered for non-linear communication lines such as the
zigzag chain and rectangular configuration which is a multichannel
line. The  geometric parameters optimizing such transfer are found.
To provide the required structure of the transferred state, where
certain elements must be zero after transfer of 0-order coherence
matrix, we use a unitary transformation of the extended receiver (we
consider a 4-qubit extended receiver for a 3-qubit receiver). These
aspects of the spin chain evolution under XXZ-Hamiltonian outline
features typical for this Hamiltonian and can be useful for  further
development of quantum information transfer in solids.

\section*{Funding} This work was funded  by Russian Federation represented by the Ministry
of Science and Higher Education of Russian Federation (grant number 075-15-2020-788).

\end{document}